\documentclass[preprint, prb, showpacs, superscriptaddress]{revtex4}
\usepackage{graphicx}
\usepackage{multirow}
\usepackage{amssymb} 
\usepackage{amsmath} 
\usepackage[utf8]{inputenc} 
\begin{document}
\topmargin-1.0cm

\title {
Maximally localized Wannier function within linear combination of pseudo-atomic orbital method: Implementation and applications to transition-metal-benzene complex
}

\author{Hongming Weng}\email[Corresponding author E-mail:]{hmweng@jaist.ac.jp}
\affiliation {Research Center for Integrated Science, Japan Advanced
Institute of Science and Technology, Nomi, Ishikawa 923-1292, Japan}
\author{Taisuke Ozaki}
\affiliation {Research Center for Integrated Science, Japan Advanced
Institute of Science and Technology, Nomi, Ishikawa 923-1292, Japan}
\author{Kiyoyuki Terakura}
\affiliation {Research Center for Integrated Science, Japan Advanced
Institute of Science and Technology, Nomi, Ishikawa 923-1292, Japan}

\date{\today}

\begin{abstract}
Construction of maximally localized Wannier functions (MLWFs) has been implemented within the linear combination of pseudo-atomic orbital (LCPAO) method. Detailed analysis using MLWFs is applied to three closely related materials, single benzene (Bz) molecule, organometallic Vanadium-Bz infinite chain, and V$_2$Bz$_{3}$ sandwich cluster. Two important results come out from the present analysis: 1) for the infinite chain, the validity of the basic assumption in the mechanism of Kanamori and Terakura for the ferromagnetic (FM) state stability is confirmed; 2) for V$_2$Bz$_3$, an important role played by the difference in the orbital energy between the edge Bzs and the middle Bz is newly revealed: the on-site energy of p$\delta$ states of edge Bzs is higher than that of middle Bz, which further reduces the FM stability of V$_2$Bz$_3$.
\end{abstract}


\maketitle

\section{introduction} \label{introduction}
The electronic ground state of a periodic system is solved in a set of Bloch functions (BFs). They are eigenfucntions of both the Hamiltonian and lattice translation operators and characterized by two good quantum numbers $n$ and $\mathbf{k}$, the band index and crystal momentum, respectively. Though BFs are widely used in electronic structure calculations, they are difficult to be visualized due to their delocalized nature and hence do not offer an intuitive physical picture for chemical bonding and other local correlations. An alternative representation which can overcome these weaknesses is Wannier functions (WFs). Compared with BFs, WFs are localized in real space and constitute a description in terms of localized functions. The localization properties of WFs depend on the choice of phase factors of the BFs. For a group of isolated bands (isolated means this set of bands are connected among themselves by degeneracies in energy, but separated from others by finite energy gaps in the whole Brillouin zone), the degrees of freedom in phase factors of BFs are equivalent to unitary transformations among themselves at each $\mathbf{k}$. Marzari and Vanderbilt\cite{MVMLWF} developed a procedure which minimizes the spread of WFs (the second moment around their centers) by refining this degree of freedom. This procedure leads to WFs that are called maximally localized Wannier functions (MLWFs). If the bands of interest are not isolated and attached to, or cross with, other bands, a prescription for extracting the interested bands out of entangled bands is required. This disentanglement procedure was proposed by Souza, Mazari and Vanderbilt \cite{Souza:2001qq} which servers as a pre-processing before refining the unitary transformations among the selected bands. Another method for constructing WFs with optimal localization properties is based on the $N$th order muffin-tin-orbital (NMTO) method.\cite{NMTOWF, d1orbitaordering} In this work, we only consider the former approach.

MLWFs have stimulated intensive interests since it brings new hope to calculate several properties of materials which are quite hard to do within the representation of BFs. Since they are real in contrast to the complex BFs and well localized in real space, one can visualize them and gain intuitive physical insight into the nature of chemical bonding. It is also possible to extract some characteristic parameters such as the MLWFs' centers and spreads. The displacements of MLWFs' center is directly related with modern theory of polarization. The hopping integrals among MLWFs from parameter-free first-principles calculations can be used to construct model Hamiltonians, or as a starting point for LDA+$U$\cite{miyake, cLDAUMLWF, UMLWF} or LDA+DMFT\cite{DMFTWF} calculations for strongly-correlated systems. Band structure interpolation based on Hamiltonian in WFs representation is quite efficient, which can be used for highly accurate integration in reciprocal space\cite{MLWFinterpolation} such as that in calculating anomalous Hall effect\cite{ahcMLWF, fermisurfaceahc} and electron-phonon coupling.\cite{epcMLWF, epcprl} MLWFs are also used for linear-scaling calculations for large systems. 

In this paper, we report briefly the implementation of MLWFs within OpenMX,\cite{openmx} a first-principles electronic structure calculation software package, which is based on the linear combination of pseudo-atomic orbital (LCPAO) basis functions and norm-conserving pseudopotentials within local density approximation (LDA) or generalized-gradient approximation (GGA). Since OpenMX is designed for large-scale {\it ab initio} calculations on parallel computers, our implementation is anticipated to allow a fast computation of MLWFs for a wide variety of materials such as biomaterials, carbon nanotubes, magnetic materials with different complex geometrical structures. 

In this work, one of the organometallic compounds vanadium-benzene sandwich-like complex,\cite{VBzMechanism, TMBzDesign} V$_n$Bz$_{n+1}$, is reexamined by our newly generated MLWFs. This complex is one of the analogues of ferrocene, a prototype of metallocene. Experimentally, V$_n$Bz$_{n+1}$ with $n\leq4$ have been found to be one-dimensional cluster and have ferromagnetic (FM) ground state with the total magnetic moment increasing nearly linearly with the cluster size.\cite{FMVBz} Inspired by these findings, several theoretical calculations\cite{fmVBzchain, xiangVBz} have been made on (VBz)$_{n=\infty}$, an ideal one-dimensional infinite chain. In those works, (VBz)$_{n=\infty}$ is found to have highly stable FM ordering and shows half-metallic behavior. Double exchange was proposed to be the mechanism of FM ordering.\cite{xiangVBz} On the other hand, by examining the electronic structure from GGA+$U$ calculations, we have proposed that the mechanism of FM stability should be that proposed by Kanamori and Terakura.\cite{KTmechanism} While compared with the infinite chain, finite V$_n$Bz$_{n+1}$ clusters are found to have much weaker FM stability though the same mechanism is applicable. By using a simple tight-binding model, we have shown that absence of p-d hybridization in one side of edge Bz leads to magnetic polarization of edge Bz even for the AFM coupling of two V atoms, which reduces the total energy of AFM situation and thus reduces the FM stability energy largely.\cite{VBzMechanism, TMBzDesign} In the present work, a much more straightforward and quantitative analysis in the representations of MLWFs shows that there is another important role of the energy difference between the edge Bz and middle Bz, which further destabilizes the FM state against AFM one. By considering this, the tight-binding model constructed now gives more consistent result with that of the direct first-principles calculation.          

In the following, we will describe technical issues regarding the construction of MLWF within the LCPAO method and then three closely related examples are studied, namely, Bz molecule, V-Bz infinite chain and V$_n$Bz$_{n+1}$ ($n$=2) finite cluster, to demonstrate the successful applications of our implementation. Finally, we will conclude in Section IV. 
 
\section{Methodology} \label{Methodology}
We will briefly introduce the theory of MLWFs. Only those aspects that are closely related with our LCPAO method will be described in detail. Details of general aspects can be found in the original papers Ref. \onlinecite{MVMLWF} and Ref. \onlinecite{Souza:2001qq}. Some other technical issues can be found in Ref. \onlinecite{wannier90}, which introduces another implementation of constructing MLWF, wannier90.\cite{wf90}

\subsection{Maximally localized Wannier functions}\label{mlwf}
The $n$th WF, $n$ being the band index, localized in unit-cell at $\mathbf R$ is defined as Fourier transforms of an isolated band expressed by BF $\psi_{n, \mathbf k}$ as follows:
\begin{equation}
|w_{n, \mathbf R} \rangle=\frac{\sqrt{N}V}{(2\pi) ^3}\int_{BZ}d{\mathbf k}e^{-i{\mathbf k}\cdot{\mathbf R}}e^{-i\phi_{n,\mathbf{k}}}|\psi_{n, \mathbf k}\rangle ,
\end{equation}
where the integral is performed over the whole Brillouin zone (BZ). $V$ is the volume of unit cell and $N$ is the number of unit cells in the sample. $e^{-i\phi_{n,\mathbf{k}}}$ is the undetermined phase factor, which brings indeterminacy of WF even transformed from a single isolated band. For a more general case with an isolated group of bands, $e^{-i\phi_{n,\mathbf{k}}}$ is generalized to a unitary transformation matrix $U^{(\mathbf{k})}$:
\begin{equation}
|w_{n, \mathbf R} \rangle=\frac{\sqrt{N}V}{(2\pi) ^3}\int_{BZ}d{\mathbf k}\sum_{m=1}^{N_{w}}U_{mn}^{(\mathbf{k})}e^{-i{\mathbf k}\cdot{\mathbf R}}|\psi_{m, \mathbf k}\rangle
\end{equation}
where $N_{w}$ is the total number of BFs in the isolated group of bands, the same as the number of WFs. In one single isolated band case, it has been proven that a suitable choice of the phase of $e^{-i\phi_{n,\mathbf{k}}}$ leads to WFs which are real and exponentially decaying in real space.\cite{realWF} In multi-band case, the arbitrariness in the gauge transformation $U^{(\mathbf k)}$ can be exploited. According to Marzari and Vanderbilt,\cite{MVMLWF} among all of the arbitrary choices, a particular set will minimize the total spread of WFs, which is defined as
\begin{equation}
\Omega[\{U^{(\mathbf k)}\}]=\sum_{n}[\langle r^{2}\rangle_n-\langle r \rangle_n^2],
\end{equation}
where $\langle r^{2}\rangle_n$ and $\langle r \rangle_n$ are the expectation values of operators $r^2$ and $\mathbf{r}$ on the $n$th WF, respectively. Both expectations expressed in WFs can be transformed into those in BFs as shown by Blount.\cite{BlountSSP} In practical calculation, a uniform k-grid is sampled to calculate the derivation of cell-periodic part of BFs in reciprocal space within finite difference approximation and the integral over k space is performed with summation over this grid. It is demonstrated that the dependence of $\Omega$ on the gauge transformation $U^{(\mathbf k)}$ is determined only by the so-called overlap integrals $M^{\mathbf{k,b}}$:
\begin{equation}
M_{mn}^{\mathbf{k,b}}=\langle \psi_{m,\mathbf{k}}|e^{-i\mathbf{b}\cdot \mathbf{r}}|\psi_{n,\mathbf{k+b}}\rangle=\langle u_{m,\mathbf{k}}(\mathbf{r})| u_{n,\mathbf{k+b}}(\mathbf{r})\rangle,
\end{equation}
where $u_{m,\mathbf{k}}(\mathbf{r})$ is the cell-periodic part of the Bloch states $\psi_{m,\mathbf{k}}=u_{m,\mathbf{k}}(\mathbf{r})e^{-i\mathbf{k}\cdot\mathbf{r}}$ and $\mathbf{b}$ is the vector connecting neighboring $\mathbf{k}$-points in the regularly discretized mesh of $\mathbf{k}$-points. $M_{mn}^{\mathbf{k,b}}$ is at the center of optimizing the spread of WFs since both the spread function itself and its gradient with respect to $U^{(\mathbf k)}$ are determined by it. Actual calculation of $M_{mn}^{\mathbf{k,b}}$ depends on the basis set used for electronic structure calculation and will be described in the following subsection.
 
\subsection{$M_{mn}^{k,b}$ in LCPAO method}\label{Mmnkb}
The wave-function within LCAPO method is defined as follow:
\begin{equation}
\psi_{m,\mathbf{k}}(\mathbf{r})=\frac{1}{\sqrt{N}}\sum_p^N e^{i\mathbf{R}_p\cdot\mathbf{k}}\sum_{i,\alpha}C_{m,i\alpha}^{(\mathbf{k})}\phi_{i\alpha}(\mathbf{r}-\mathbf{\tau}_i-\mathbf{R}_p)
\end{equation}
where $\phi_{i\alpha}(\mathbf{r}-\mathbf{\tau}_i-\mathbf{R}_p)$ is the pseudo-atomic orbital $\alpha$ centered on site $\mathbf{\tau}_i$ in unit-cell $\mathbf{R}_p$ and $C_{m,i\alpha}^{(\mathbf{k})}$ is the linear combination coefficients of them at $\mathbf{k}$ for band $m$. The overlap integral matrix element $M_{mn}^{\mathbf{k,b}}$ is
\begin{equation}
\begin{aligned}
M_{mn}^{\mathbf{k,b}} ={}&\langle u_{m,\mathbf{k}}(\mathbf{r})| u_{n,\mathbf{k+b}}(\mathbf{r})\rangle \\
 ={}&\langle \psi_{m,\mathbf{k}}|e^{i\mathbf{k}\cdot\mathbf{r}}e^{-i(\mathbf{k+b})\cdot\mathbf{r}}|\psi_{n,\mathbf{k+b}}\rangle \\
 ={}&\frac{1}{N}\sum_{p,q}^{N}e^{-i(\mathbf{R}_p-\mathbf{R}_q)\cdot\mathbf{k}}\sum_{i\alpha,j\beta}{C_{m,i\alpha}^{(\mathbf{k})}}^{\ast}C_{n,j\beta}^{(\mathbf{k+b})}\times \\
 {}& \langle \phi_{i\alpha}(\mathbf{r}-\mathbf{\tau}_i-\mathbf{R}_p)|e^{-i(\mathbf{r}-\mathbf{R}_q)\cdot\mathbf{b}}|\phi_{j\beta}(\mathbf{r}-\mathbf{\tau}_j-\mathbf{R}_q)\rangle
 \end{aligned}
\end{equation}
Defining that $\mathbf{r}'=\mathbf{r}-\mathbf{\tau}_i-\mathbf{R}_p$, it becomes
\begin{equation}
\begin{aligned}
M_{mn}^{\mathbf{k,b}} ={}&\frac{1}{N}\sum_{p,q}^{N}e^{-i\mathbf{k}\cdot(\mathbf{R}_p-\mathbf{R}_q)}\sum_{i\alpha,j\beta}{C_{m,i\alpha}^{(\mathbf{k})}}^{\ast}C_{n,j\beta}^{(\mathbf{k+b})} \times \\
{}&  \langle \phi_{i\alpha}(\mathbf{r}')|e^{-i(\mathbf{r}'+\mathbf{\tau}_i+\mathbf{R}_p-\mathbf{R}_q)\cdot\mathbf{b}}|\phi_{j\beta}(\mathbf{r}'+\mathbf{\tau}_i-\mathbf{\tau}_j+\mathbf{R}_p-\mathbf{R}_q)\rangle,
\end{aligned}
\end{equation}
in which each term depends on only the relative position $\mathbf{R}_p-\mathbf{R}_q$. Therefore, Eq. (7) can be written as
\begin{equation}
\begin{aligned}
M_{mn}^{\mathbf{k,b}}={}&\sum_{p}^{N}e^{-i\mathbf{k}\cdot\mathbf{R}_p}\sum_{i\alpha,j\beta}{C_{m,i\alpha}^{(\mathbf{k})}}^{\ast}C_{n,j\beta}^{(\mathbf{k+b})}e^{-i\mathbf{b}\cdot(\mathbf{\tau}_i+\mathbf{R}_p)} \times \\
{}&  \langle \phi_{i\alpha}(\mathbf{r}')|e^{-i\mathbf{r}'\cdot\mathbf{b}}|\phi_{j\beta}(\mathbf{r}'+\mathbf{\tau}_i-\mathbf{\tau}_j+\mathbf{R}_p)\rangle.
\end{aligned}
\end{equation}
A uniform grid in real space is used to perform the integral. In practice, $e^{-i\mathbf{r}'\cdot\mathbf{b}}$ is expanded in terms of $\mathbf{r}'\cdot\mathbf{b}$
\begin{equation}
e^{-i\mathbf{r}'\cdot\mathbf{b}}=1-i\sum_{i=1}^{3}r'_i b_i+\frac{1}{2!}\sum_{i,j=1}^{3}\frac{\partial^2(-i\mathbf{r}'\cdot\mathbf{b})}{\partial x_i \partial x_j}x_i x_j+ ...
\end{equation}
Since $\mathbf{r}'$ is an extended operator, we find that an expansion up to 4th order is needed to well conserve the unitary condition of $M_{mn}^{\mathbf{k,b}}$. A denser $k$-space sampling will give smaller $\mathbf{b}$, which helps to improve this conservation and lower order expansion can be used. 
   
\subsection{Initial guess of MLWFs}\label{guess}
The minimization of the spread function begins with an initial guess for the target WFs. Following the approach proposed by Marzari and Vanderbilt, a set of trial functions $|g_n(\mathbf{r})\rangle$, $n \in [1, N_w]$, are taken as the initial guess of the $N_w$ target MLWFs. In our LCPAO method, it is very convenient and natural to take the localized pseudo-atomic orbitals, which are the bases for expanding BFs, as the initial guesses. The center of each orbital can be put at any place in the unit-cell and all the other characters for atomic orbital, such as the radial and angular functions, can be easily controlled by using suitable pseudo-atomic orbitals. The possible hybrids among the atomic orbitals are also available. 

\subsection{General settings in OpenMX calculation}\label{setting}
Three closely related materials, single Bz molecule, V-Bz infinite chain and V$_2$Bz$_3$ cluster, are chosen to demonstrate our implementation. To calculate their electronic structures, PAOs are generated by a confinement potential scheme.\cite{pao} For both hydrogen and carbon, the cutoff radius is 5.0 a.u. while it is 6.5 a.u. for vanadium. When generating pseudopotential, the semicore 3s and 3p states of V atom are included as valence states. The exchange correlation energy functional within GGA\cite{pbe} is used for all the systems. Double-valence and polarization orbitals of each element are included as basis set: s2p2, s2p2d1 and s2p2d2f1 are used for H, C and V, respectively.\cite{pao} In the electronic structure calculation, the real-space grid technique\cite{realgrid} is used with an energy cutoff of 250 Ry in numerical integrations and in the solution of the Poisson equation. The GGA+$U$ calculation is done with the approach proposed in Ref. \onlinecite{plusU}. The geometrical structure of these materials are relaxed until the forces are less than $1.0\times10^{-4}$ a.u. For molecular or cluster calculation, a supercell is used and the size is as large as 17 \AA$ $ to assure that the interaction between neighboring cells can be neglected. 

\section{results and discussions} \label{result}
\subsection{Benzene Molecule}\label{C6H6}
Benzene (Bz) molecule is firstly studied with only $\Gamma$-point sampled in BZ. Nine molecular orbitals (MO) around the HOMO and LUMO states are shown in Fig. 1 together with their eigenvalues and symmetries. Clearly HOMO-2, LUMO+2 and doubly degenerate HOMO, LUMO are composed of six $p_z$ orbitals on carbon atoms. All of the nine MOs are used to construct six MLWFs, i. e.,  $N_{win}$=9  (Note that $N_{win}$ denotes the number of band branches within the selected energy window as defined in Ref. \onlinecite{Souza:2001qq}.) and $N_w$=6. A physically intuitive initial guess for this set of target MLWFs are six $p_z$ orbitals on each carbon atom. The disentangling process proposed in Ref. \onlinecite{Souza:2001qq} is used to select an optimized $6\times6$ subspace, which minimizes the gauge invariant part of the spread function, $\Omega_{I}$. After that, a steepest-decent (SD) method is used to minimize the gauge-dependent part of the spread function to find the proper gauge transformation. With this initial guess, the total spread converged to $10^{-13}$ \AA$^2$ within 50 SD steps. The obtained six MLWFs are obviously identical to each other and have the spread of 0.943 \AA$^2$, which is only slightly smaller than the initial spread of 0.944 \AA$^2$. MLWFs are real and have similar shape to the atomic $p_z$ orbital as shown in Fig. 2. Although the initial guess of MLWFs are centered on each carbon atom, the converged MLWFs' centers are slightly shifted outward the gravity center of Bz by about 0.07 \AA$ $ along each nearby C-H bond direction. 

In Table. I, we listed the hopping integrals between these six MLWFs in the same unit cell. $w_i$ means the MLWF centered around carbon atom i as shown in Fig. 2. The first column shows the orbital energy of the $p_z$ type MLWFs. The hopping integral from $w_1$ to its two nearest neighbors $w_2$ and $w_6$ are -2.88 eV and those to further neighbors $w_3$, $w_5$ and $w_4$ are 0.19 and -0.24 eV, respectively. The sign change in the hopping integral as the distance increases suggests that the WFs have oscillating tails to satisfy orthogonality relation.
\begin{figure}
\centering
\includegraphics[scale=0.8]{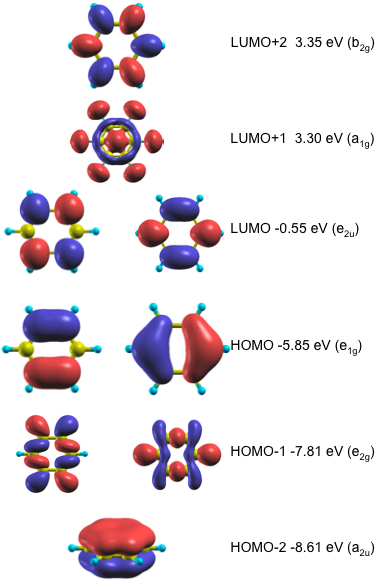}
\caption{(color online). Molecular orbitals of Bz around the HOMO and LUMO states. The energy eigenvalue and symmetry of each orbital are shown. Doubly degenerate orbitals are shown together on the same level. }\label{fig1}
\end{figure}

\begin{figure}
\centering
\includegraphics[scale=0.5]{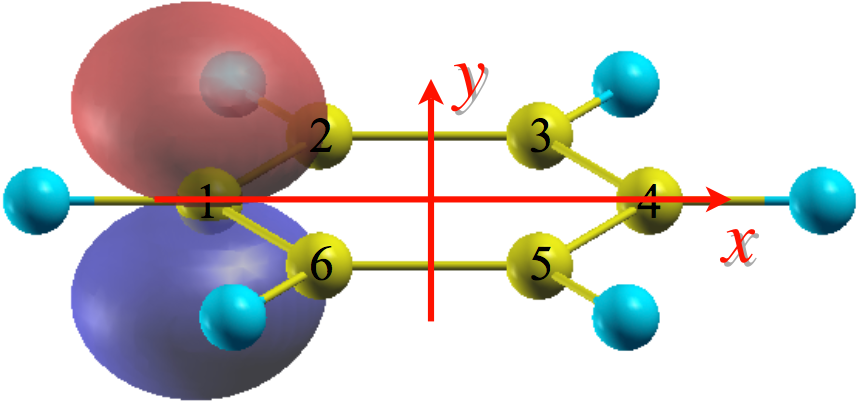}
\caption{(color online). The MLWF around the carbon atom 1 obtained for a Bz molecule is plotted with isovalue=0.1. All the six MLWFs are identical and their centers are shifted from nearby C atomic sites along each C-H bond by about 0.07 \AA.}\label{fig2}
\end{figure}
 
\begin{table}
\caption{The first column shows the orbital energy of the $p_z$ like MLWF of a Bz molecule as an isolated one (first row). The rest of the columns are for the hopping integrals from $w_1$ to other five MLWFs. In the second row is for the  V-Bz infinite chain. The numbers outside (inside) of the parenthesis are for spin up (spin down) channels. Energies are in eV.}\label{tab1}
\begin{tabular}{c|c|ccc}
\hline \hline
\multirow{2}{*}{} & orbital energy & \multicolumn{3}{c}{hopping integrals between $w_1$ and $w_i$ ($i=2\sim 6$)}\\
\cline{2-5}
     & $w_1$  & $w_2$,$w_6$ & $w_3$,$w_5$  &$w_4$ \\
\hline
Bz &  -3.01   &   -2.88   &   0.19  &   -0.24   \\
  \hline
(VBz)$_{\infty}$ & -4.23(-4.33) & -2.64(-2.61) & 0.08(0.08) &  -0.20(-0.19) \\
\hline \hline
\end{tabular}
\end{table}

\subsection{Ideal infinite (VBz)$_{n=\infty}$ chain}\label{VBz}
In our previous work,\cite{VBzMechanism, TMBzDesign} we have analyzed the mechanism of ferromagnetism stability in V-Bz infinite chain. Here we reexamine this system by using MLWFs. The spin polarized band structure FM V-Bz chain is shown in Fig. 3. Due to the symmetry, when V is sandwiched with Bz, its 3d orbitals will hybridize with Bz's HOMO-2, HOMO, LUMO to form 3 types of bonds, namely $\sigma$, $\pi$ and $\delta$. Although HOMO-1 has also the $\delta$ symmetry, its hybridization with d$_{xy}$ and d$_{x^2-y^2}$ orbitals of V is negligible because the wave function of HOMO-1 is confined within Bz molecular plane. The 4s state of V is pushed to a high energy by about 5-6 eV with strong hybridization with HOMO-2. The original 4s electrons of V are transfered to the states near the Fermi level in Fig. 3. By including LUMO+2 orbital, we construct eleven MLWFs (six Bz MOs and five V 3d-orbitals) from the eigenstates in the outer window from -8.4 eV to 6 eV as shown in Fig. 3. To demonstrate our implementation that can be applied to solid system, a $2\times2\times20$ $k$-grid is used. As in the Bz molecule case, the disentangling process is also necessary. We started from the initial guess of six $p_z$ orbitals on each carbon atom and five d orbitals on V atom. The tolerance for convergence is $10^{-12}$ \AA$^2$ for both gauge-invariant and -dependent parts of spread functions. Two schemes are used for optimizing the spread function. The SD method is used for first hundreds steps and it converged to $10^{-3}$ \AA$^2$. After that conjugate gradient (CG) method is adopted to continue the minimization, which takes about 40 steps to converge to $10^{-12}$ \AA$^2$. This hybrid scheme are found to be stabler than just using CG method in some cases, while faster than just using SD method.

The quality of the obtained MLWFs can be seen from the interpolated band structure obtained by the band parameters for MLWFs shown in Fig. 3, which overlaps that of the original band at most $\mathbf{k}$ points. The obvious discrepancy happens in the $\sigma$ bands around the Fermi level originally composed of $d\sigma$ and LUMO+1 states. This is because LUMO+1 state is discarded after disentangling. However, these two states have decreasing hybridization strength from $\Gamma$ to X points, which leads to better reproduction of bands near the X point.\cite{noteband} The obtained MLWFs fulfill the requirement of real valuedness.\cite{realMLWF} They are six $p_z$ type orbitals centered nearby each carbon atom with a shift about 0.05 \AA $ $ along the C-H bond and five 3d-like orbitals centering on V atom site. The spreads for each MLWF and gauge decomposition are listed in Table II. The spreads of five d orbitals are classified into three categories corresponding to three types of bonds with Bz. The spread of d$\pi$ like MLWF is larger than that of d$\delta$ is due to the stronger hybridization with Bz's $p_z$ like MLWFs while the largest spread of d$\sigma$ like MLWF is due to the d$\sigma$-d$\sigma$ overlap along the chain. It is also noticed that the spreads of d orbitals in majority (up) spin channel are smaller than those in minority (down) one. As Bz has negative spin polarization against V, the spreads of spin down are also smaller than that of spin up, implying the occupied states may be more localized than unoccupied ones.

\begin{figure}
\centering
\includegraphics[width=0.6\textwidth]{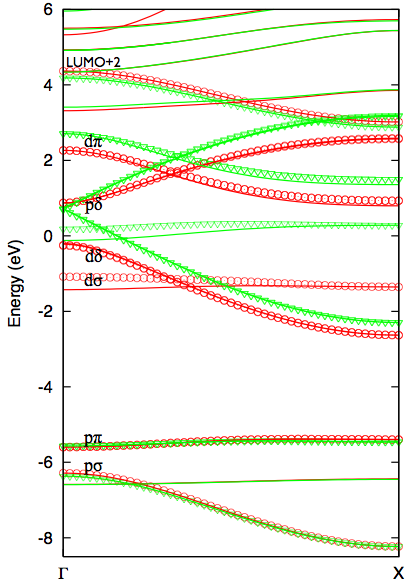}
\caption{(Color online) Band structure of FM V-Bz infinite chain. Lines are from ordinary first-principles calculations while symbols are those from Wannier interpolation. Red color is for spin up while green is for spin down. The component of interpolated bands at $\Gamma$ point are labelled, too.}\label{fig3}
\end{figure}

\begin{table}
\caption{The spreads (in \AA$^2$) of MLWFs of V-Bz infinite chain for GGA and GGA+$U$ ($U$=3.0 eV) calculations. $p_z$ means the $p_z$ type MWLF, so as d$_{z^2}$, d$_{x^2-y^2}$, etc. $\Omega_I$ is the gauge invariant part of the spread function. $\Omega_D$ and $\Omega_{OD}$ represent the diagonal and off-diagonal contributions to the gauge dependent part of the spread function, respectively.}\label{tab2}
\begin{tabular}{c|c|c|c|c|c|c|c}
\hline \hline
  spin & $p_z$ on Bz & d$_{z^2}$ & d$_{x^2-y^2}$,d$_{xy}$  & d$_{xz}$,d$_{yz}$ & $\Omega_I$ &$\Omega_D$ &$\Omega_{OD}$\\
\hline
  up(GGA)   &  1.209   &   1.155  &   0.813    &   1.052  & 11.951 & 0.001 & 0.188 \\
\hline
  down(GGA)   &   1.186   &  1.504  &  0.870    &   1.088 & 12.264 & 0.001 &  0.273 \\ 
\hline 
  up(+$U$)   &  1.244   &   1.080  &   0.781    &   1.096  &  12.158 & 0.001  &  0.140 \\
\hline
  down($+U$)   &   1.179   &  1.762  &  0.935    &   1.221 &  12.814 &  0.001 & 0.341  \\ 
 \hline \hline
\end{tabular}
\end{table}

The hopping integrals between these MLWFs can give more detailed information on bonding nature and physical mechanism in this system. The hopping integrals between six $p_z$ orbitals are compared with those in an isolated Bz molecule in Table I. The on-site energy becomes -4.23 eV and -4.33 eV for spin up an spin down, respectively, which are lower than the one in an isolated Bz molecule due to the crystal field formed by V$^{2+}$ ions. The hopping integrals between neighboring $p_z$ orbitals in one carbon ring are a little smaller than those in Bz molecule since the optimized C-C bond length in V-Bz chain is slightly larger than that in an isolated molecule.\cite{VBzMechanism} The on-site energy of five d orbitals listed in Table III are also classified into three types corresponding to three types of bonds. The hopping integrals between d and $p_z$ orbitals depend little on the spin polarization. Here only the hopping integrals between $w_1$ and d orbitals are listed. Those for other $p_z$ like MLWFs can be obtained by appropriate rotation of d orbitals. 
The $p_z$-$d$ hopping integrals decay quickly along the $c$-axis, the chain direction. For example, the $p_z$-$d$ hopping integrals for the next nearest neighboring Bz are about 2 orders of magnitude smaller than those for the nearest neighbors and those for the third nearest neighbor can be ignored. This quick decaying property also indicates the good localization of obtained MLWFs.


\begin{table}
\caption{
The onsite energies (second row) of d-type MLWFs and the $p_z$-$d$ hopping integrals (third row) between one of the $p_z$-type MLWF $w_1$ and d-type ones in FM V-Bz chain. In the fourth row, the $p_z$-$d$ hopping integrals with $p_z$ converted to molecular orbitals with $\sigma$, $\pi$ and $\delta$ symmetries. The values outside (inside) the parentheses are spin up (spin down) channel and in unit of eV. The results are from the GGA calculations.
}\label{tab3}
\begin{tabular}{c|c|c|c|c|c}
\hline \hline
                &    d$_{z^2}$     &     d$_{x^2-y^2}$    &     d$_{xy}$     &    d$_{xz}$    &    d$_{yz}$    \\
\hline
 on-site     &  -3.47(-1.99)     &   -2.33(-1.35)     &  -2.33(-1.35)     &  -1.78(-1.21)   & -1.78(-1.21)  \\
 \hline
 $w_1$     &  0.01(0.07)        &   -0.77(-0.78)      &   0.00(0.00)      &   -0.96(-0.97)   &   0.00(0.00)   \\
 \hline
 $MO$      &  0.026(0.171)    &  -1.34(-1.36)       &  1.34(1.36)       &   -1.66(-1.68)   &   1.66(1.68)  \\
\hline \hline
\end{tabular}
\end{table}

 
Similar calculation with $U$=3.0 eV is performed to study how $U$ influences the MLWFs. Compared with those in GGA calculation, +$U$ makes the d-type ($p_z$-type) MLWFs with majority spin more localized (extended), while those with minority spin more extended (localized). The on-site energy of $p_z$ orbitals becomes -3.97 and -4.46 eV for spin up and spin down channels, respectively. The enhanced spin splitting of $p_z$ orbitals is induced by the enhanced spin splitting in V d orbitals. The hopping integrals between different $p_z$ orbitals keep nearly the same as those in GGA case. Similarly for V d orbitals, the spin splitting of the on-site energy is enhanced. 
For example for $U$=3.0eV, the spin up (spin down) on-site energies of d$\sigma$, d$\delta$ and d$\pi$ states are -6.00 (-0.85), -3.91 (-0.37) and -1.38 (-0.27) eV, respectively. However, the $p_z$-$d$ hopping integrals have nearly no change. 
  
By using the MLWFs obtained so far, we reexamine quantitatively the tight-binding model of Ref. \onlinecite{VBzMechanism} and \onlinecite{TMBzDesign}. For this purpose, the atomic like MLWFs, $w_i$ ($i=$1 to 6), are converted to MOs with symmetries of $\sigma$, $\pi$ and $\delta$. The essential point in Ref. \onlinecite{VBzMechanism} are summarized below. HOMOs of Bz and d$_{xz}$, d$_{yz}$ orbitals having $\pi$ symmetry form quite strong bonds to hold the sandwich-like geometrical structure, while HOMO-2 and d$_{z^2}$ orbitals having $\sigma$ symmetry couples weakly. The orbital with strong d$_{z^2}$ character is singly occupied and plays the role as a trigger of spin splitting. LUMOs and d$_{x^2-y^2}$, d$_{xy}$ orbitals having $\delta$ symmetry form bonds to mediate magnetic coupling and they are mainly responsible for the stability of FM states. Our proposed mechanism for FM stability is described in Fig. 4 following Kanamori and Terakura.\cite{KTmechanism} In this picture, the essential assumption is that the on-site energy of p$\delta$ states (LUMOs of Bz) should be between those of the spin majority and minority d states of $\delta$ symmetry before switching on the p-d hybridization. Then, p$\delta$ states could be negatively spin polarized due to the p-d hybridization leading to stability of the FM configuration. The validity of the above assumption can be easily checked by the molecular orbitals formed with $w_i$ ($i=$1 to 6). 

\begin{figure}
\centering
\includegraphics[width=\textwidth]{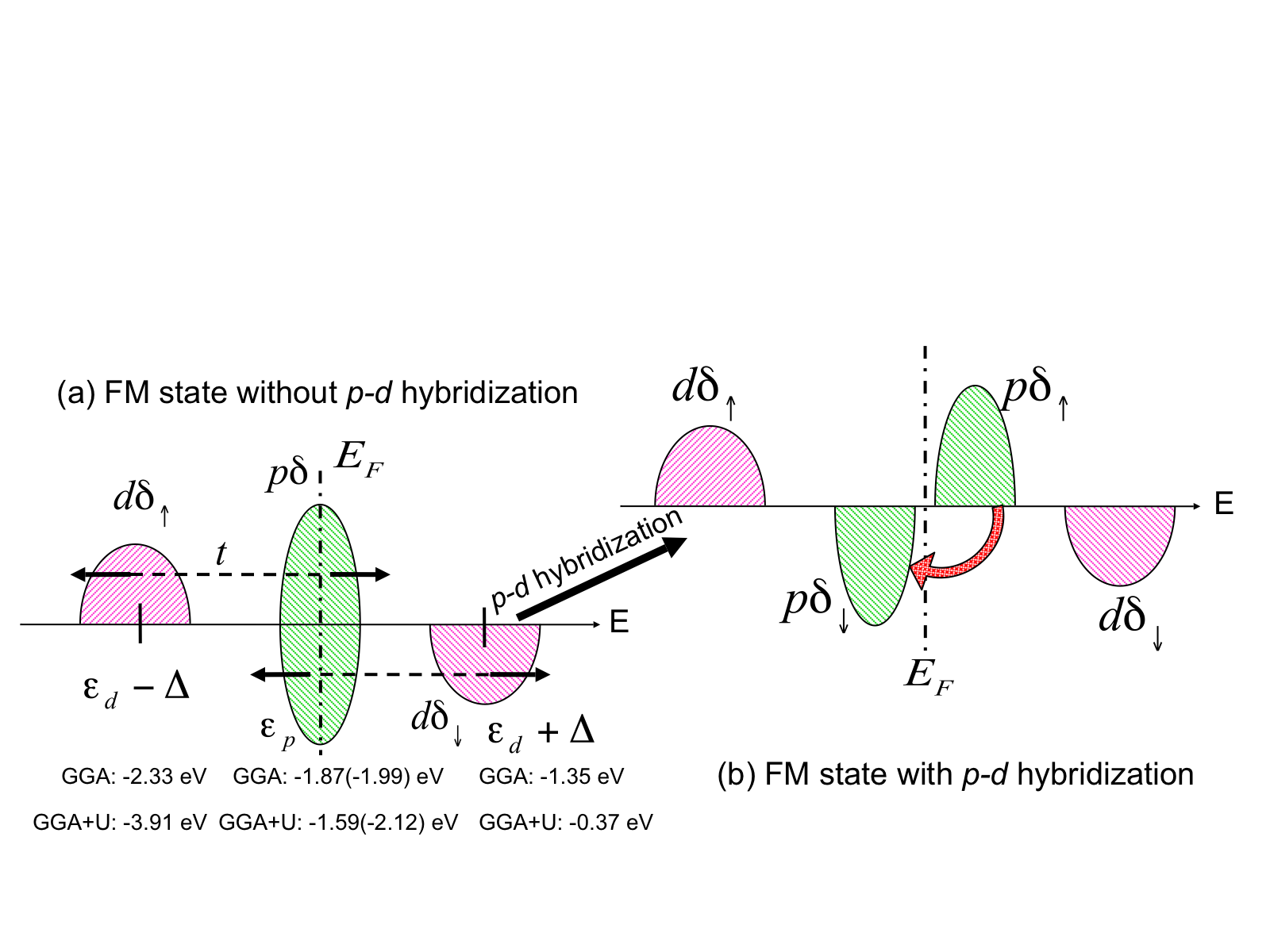}
\caption{(Color online) Schematic pictures describing the mechanism in which the $p-d$ hybridization stabilizes the FM states. The DOS (a) without and (b) with $p-d$ hybridization is plotted. The energy gain due to the charge transfer associated with the magnetic relaxation of p$\delta$ states is indicated. Before $p-d$ hybridization, the on-site energies of p$\delta$ and d$\delta$ states are given in eV for both GGA and GGA+$U$ calculations. See the text for more details.}\label{fig4}
\end{figure}

Similarly to a single Bz molecule case, diagonalizing $6\times6$ subspace Hamiltonian constructed from the orbital energy and hopping integrals listed in Table I gives the eigen-energies and eigenstates of HOMO-2, HOMOs, LUMOs and LUMO+2. LUMOs, which play crucial roles in the FM stability have eigen-energies of -1.87 and -1.99 eV for spin up and spin down case, respectively. These two values are just sitting in-between the on-site energies of majority d$\delta$ state, -2.33 eV and minority one, -1.35 eV in GGA. This picture is valid also in GGA+$U$ calculation. In this case the spin up and spin down p$\delta$ states are located at -1.59 and -2.12 eV, respectively, which are well sitting between spin up d$\delta$ states at -3.91 eV and those of spin down at -0.37 eV. The p-d hopping integral between LUMOs and d$\delta$ (d$_{x^2-y^2}$, d$_{xy}$) orbitals can be easily obtained by using the hopping integrals between the atomic-like $p_z$ MLWFs and d$\delta$ orbitals. For example, one of the LUMO states is given by\begin{equation}
|LUMO\rangle=\frac{1}{2}(|w_2\rangle-|w_3\rangle+|w_5\rangle-|w_6\rangle).
\end{equation}
Then in spin up case by using the values listed in Table III the hopping integral from this LUMO to d$_{xy}$ state is
\begin{equation}
\langle d_{xy}|\hat{H}|LUMO\rangle = 1.34.
\end{equation}
Similarly, the hopping integrals between HOMOs and d$\pi$ orbitals and that between HOMO-2 and d$\sigma$ orbital can also be calculated as listed in Table III. The bonding strength of different bonds can be seen from these hopping integrals. For $\sigma$ and $\pi$ bonds, the values are around 0.026(0.171) and 1.66(1.68) eV, respectively. The larger difference between spin up and spin down values in $\sigma$ bond case is a reflect of larger difference in spreads of d$_{z^2}$ like WF in spin up and down channels. Therefore, by the analysis using MLWFs, the physical pictures proposed in our previous work\cite{VBzMechanism} are directly and quantitively confirmed. 


\subsection{Finite V$_2$Bz$_3$ cluster}\label{V2Bz3}
We discussed the difference between infinite V-Bz chain and finite V$_n$Bz$_{n+1}$ clusters in our previous two works.\cite{VBzMechanism, TMBzDesign} The reasons for nearly degenerate FM and AFM states in the finite V$_2$Bz$_3$ are extensively discussed. The importance of edge Bz is emphasized when compared with the infinite V-Bz chain. In the finite cluster, the edge Bzs having p-d hybridization only with one side of V atom can have magnetic relaxation even in the AFM configuration, which is not possible in the infinite chain. This makes the AFM state stabler and reduces the relative stability of FM state against AFM state. The p-d hybridization between V and edge Bz is stronger than that between V and middle Bz, which decreases the FM stability energy further. However, the present more quantitative analysis using MLWFs has revealed an additional source for making the FM and AFM states nearly degenerate in energy. 

The MLWFs for both FM and AFM V$_2$Bz$_3$ clusters have been constructed in the same way as above. We have found that MLWFs of edge Bz are different from those of middle Bz in several ways. 1) The spreads of MLWFs from middle Bz are slightly larger than those from edge Bz by about 0.015 \AA$^2$. 2) The hopping integrals from middle Bz to V d orbitals are slightly smaller than those from edge Bzs since V is closer to edge Bzs. This is consistent with our former analysis. 3) The on-site energy of LUMOs (p$\delta$ states) on edge Bz is about 0.78 and 0.88 eV higher than those on middle Bz for spin up and spin down states, respectively. The reason for the relatively higher p$\delta$ states on edge Bz is easily understood. Since the middle Bz is sandwiched by two V$^{2+}$ ions while each of the edge Bzs have only one nearest V$^{2+}$ ion, the attractive electrostatic field on middle Bz is stronger than that on edge Bz and brings higher on-site energies for states on edge Bzs. While in our former work,\cite{VBzMechanism} we assumed that the on-site energies of p$\delta$ states are the same for both middle and edge Bzs. It is easy to check how the FM stability energy depends on the on-site energy difference between the edge Bzs and middle Bz by using the tight-binding model proposed in Ref. \onlinecite{VBzMechanism}. In that model, only $\delta$ bonds which are critical to the coupling of magnetic moments are considered. The parameters included in this model can be obtained directly from the construction of MLWFs. The on-site energy of p$\delta$ states on middle Bz, $\varepsilon_p$, is found to be -2.34 eV (average value of those in spin up and spin down channels). $\varepsilon_d$ (on site energy of non spin polarized d$\delta$ state) and $\Delta$ (exchange parameter for d$\delta$ states) can be deduced from the orbital energy of spin polarized d$\delta$ orbitals. They are found to be -2.13 and 1.18 eV, respectively. Parameters for describing hybridizations between d$\delta$ orbitals and p$\delta$ states on edge Bzs and middle Bz are $t'$ and $t$ with the values of 1.44 and 1.28 eV, respectively. By using these parameters, it is shown in Fig. 5 that the FM stability energy decreases as the energy difference in p$\delta$ states on edge Bzs and middle Bz increases. AFM state becomes stabler than FM state (FM stability energy becomes negative) when p$\delta$ states on edge Bz is about 0.67 eV higher than those on middle Bz. This critical value is a little bit different from the value, around 0.83 eV, obtained directly from MLWFs since in this tight-binding model other contributions, such as those from $\sigma$ bonds, are neglected. Therefore, we think that the difference between edge Bzs and middle Bz in p$\delta$ states is another important source for the much reduced FM stability in finite clusters. 

\begin{figure}
\centering
\includegraphics[width=\textwidth]{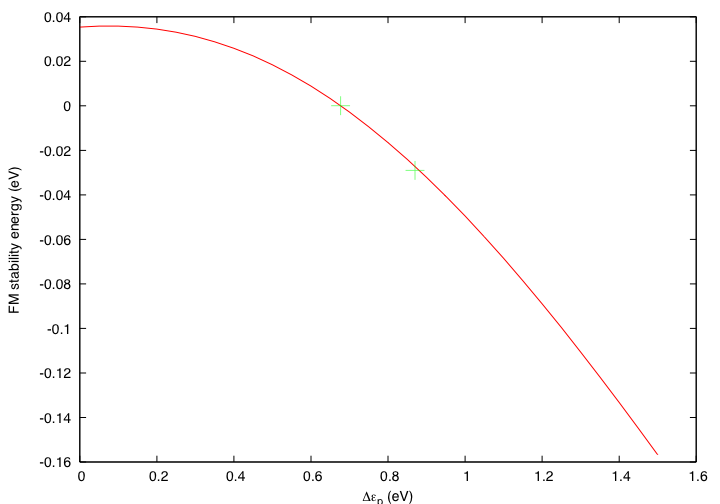}
\caption{(Color online) FM stability energy (total energy of AFM state minus that of FM state, in eV) depends on the energy difference in p$\delta$ states ($\Delta\varepsilon_p$) on edge Bzs and middle Bz. The points where the FM stability energy equals 0.0 eV and the on-site energy difference being 0.83 eV are indicated by "+" symbols. }\label{fig5}
\end{figure}

\section{Conclusion} \label{Conclusion}
We have implemented the construction of maximally localized Wannier funcitons in the formalism of linear combination of pseudo-atomic orbitals first-principles calculations. The implementation is demonstrated to be applicable to both solid and molecular systems. The disentangling procedure works well for metallic cases also. The analysis of hopping integrals obtained from MLWFs for V-Bz organometallic complex indicates that it is a proper way to find the tight-binding parameters from parameter-free {\it ab initio} calculations. MLWFs provide useful information for the study of the bonding nature and physical mechanism in materials. As demonstrated in V-Bz complexes, a new physical origin, the role of orbital energy difference between edge Bz and middle Bz neglected before, is naturally revealed by analysis with MLWFs. 

\begin{acknowledgments}
The authors thank the staffs of the Center for Information Science in JAIST and Information Initiative Center in Hokkaido University for their support and the use of their supercomputing facilities. H. M. Weng acknowledges the Research Promoting Expense for Assistant professors in JAIST. This work is partly supported by the Next Generation Supercomputing Project, Nanoscience Program and also by the Grants-in-Aid for Scientific Research in Priority Area "Anomalous Quantum Materials" both from the Ministry of Education, Culture, Sports, Science and Technology, Japan. One of the authors, T. O., is also partly supported by CREST-JST.
\end{acknowledgments}




\begin{thebibliography}{100}
\bibitem{noteband}
Another discrepancy exists in the d$\pi$ band near X point instead of $\Gamma$ point in d$\sigma$ band due to the similar reason. d$\pi$ orbitals have increased hybridization strength with another unoccupied MO close to LUMO+2 from $\Gamma$ to X, while this unoccupied MO is discarded during disentangling.
\end{thebibliography}
\end{document}